\documentclass[onecolumn,11pt]{revtex4}

\usepackage{graphicx}
\usepackage{dcolumn}
\usepackage{bm}

\input epsf

\begin{document}

\title{\Large The General Class of Accelerating, Rotating and Charged Plebanski-Demianski Black Holes as Heat Engine}

\author{\bf Ujjal Debnath\footnote{ujjaldebnath@gmail.com}}

\affiliation{Department of Mathematics, Indian Institute of
Engineering Science and Technology, Shibpur, Howrah-711 103,
India.\\}

\date{\today}

\begin{abstract}
We first review the general class of accelerating, rotating and
charged Plebanski-Demianski (PD) black holes in presence of
cosmological constant, which includes the Kerr-Newman rotating
black hole and the Taub-NUT spacetime. We assume that the
thermodynamical pressure may be described by the negative
cosmological constant and so the black hole represents anti-de
Sitter (AdS) PD black hole. The thermodynamic quantities like
surface area, entropy, volume, temperature, Gibb's and Helmholtz's
free energies of the AdS PD black hole are obtained due to the
thermodynamic system. Next we find the critical point and
corresponding critical pressure, critical temperature and critical
volume for AdS PD black hole. Due to the study of specific heat
capacity, we obtain ${\cal C}_{V}=0$ and ${\cal C}_{P}\ge 0$. From
this result, we conclude that the AdS PD black hole may be stable.
We examine the Joule-Thomson expansion of PD black hole and by
evaluating the sign of Joule-Thomson coefficient $\mu$, we
determine the heating and cooling nature of PD black hole. Putting
$\mu=0$, we find the inversion temperature. Next we study the heat
engine for AdS PD black hole. In Carnot cycle, we obtain the work
done and its maximum efficiency. Also we describe the work done
and its efficiency for a new engine. Finally, we analyze the
efficiency for the Rankine cycle in PD black hole heat engine.
\end{abstract}

\maketitle

\newpage

\sloppy \tableofcontents

\section{Introduction}

The black hole thermodynamics has become an important topic of
intensive research since Hawking's radiation of black holes
\cite{Haw1,Haw2} and considered as a crucial topic to gaining
insight into the quantum nature of gravity. From the early
discoveries that black hole area and surface gravity behave as
thermodynamic entropy \cite{B,Bek2} and temperature \cite{Haw1}
respectively. Gibbons and Hawking \cite{Gibb} have studied the
physics of anti-de Sitter (AdS) black hole due to AdS/CFT
correspondence. Hawking and Page \cite{Hawking} have studied the
thermodynamic properties of static Schwarzschild-AdS black hole.
After few years, Chamblin et al \cite{Cham1,Cham2} have
investigated the physical properties of charged
Reissner-Nordstrom-AdS black hole. If one considers charge and/or
rotation of the AdS black hole, the nature of the AdS black hole
is similar to the Van der Waals fluid \cite{Cv,Niu}. In the black
hole chemistry \cite{K1,K2}, the negative cosmological constant
($\Lambda<0$) is considered as a thermodynamic pressure
$P=-\frac{\Lambda}{8\pi}=\frac{3}{8\pi \ell^{2}}$ ($\ell$ is the
length of AdS black hole)
\cite{K3,Cre,Cald,K4,K5,K6,K7,Kubiz,Gun,K8}, has recently started
to attract a growing deal of interest. In the thermodynamic
system, the first law of black hole thermodynamics gives $\delta
M=T\delta S+V\delta P+...$ with the black hole thermodynamic
volume $V=\left(\frac{\partial M}{\partial P}\right)_{S,...}~, $
where, $M$ is the mass, $S$ is the entropy and $T$ is the
temperature of the AdS black hole. The geometry of AdS black hole
thermodynamics has been extensively studied by several authors
\cite{G1,G2,G3,G4,G5,G6,G7,G8,G9,G10,G11,G12,G13,G14,G15,G16,G17}.\\

In the context of black hole chemistry, the concept of holographic
heat engine has been proposed by Johnson \cite{John} for AdS black
hole, where he has considered the cosmological constant as a
thermodynamic variable. Based on the holographic heat engine
proposal, Setare et al \cite{Set1} have studied polytropic black
hole as a heat engine. Subsequently, Johnson \cite{Joh,Joh1,Joh2}
has analyzed the heat engine phenomena for the Gauss-Bonnet black
holes and Born-Infeld AdS black holes. Holographic heat engines
for different types of black holes have been studied in
\cite{Henn,Mo1,Fang,Rosso,J,Hhu,Ahd,Gha,Johns11}. Zhang et al
\cite{Zhan} have studied the $f(R)$ black holes as heat engines.
Heat engines of AdS black hole have been analyzed in the
frameworks of massive gravity \cite{Mo1,Hendi,Fern,Nam1},
gravity's rainbow \cite{Pana} and conformal gravity \cite{Hu1}.
Heat engine efficiency has been studied for Hayward \cite{GuoS}
and Bardeen \cite{Raj0,Run} black holes. Heat engine in three
dimensional BTZ black hole has been obtained in \cite{MoL,Bala}.
Heat engine for dilatonic Born-Infeld black hole has been analyzed
in \cite{Chan}. For charge rotating dyonic black hole, the
thermodynamic efficiency has been studied in \cite{Sadeg}. Till
now, several authors also have studied the heat engine phenomena
of black holes in various occasions
\cite{Chakra,Hen,Liu,Johns,Wei1,Avik,
Moo,Graca,YeC,Santo,Debnath1,Js,Por}.\\

Using the standard black hole thermodynamics, it was found that
the Hawking temperature of accelerating black holes is more than
Unruh temperature of the accelerated frame. Thermodynamics nature
of accelerating black holes have been discussed in
\cite{App,Asto,Anab,Abba,Han,Guha}. The thermodynamics properties
of accelerating and rotating black holes have been investigated in
\cite{Bilal,Rizw}. Also charged accelerating black hole
thermodynamics have been analyzed in \cite{Jaf,Pra1,Tavak}.
Charged rotating and accelerating black hole thermodynamics have
been studied in \cite{Shar,Anab1}. The entropy bound of horizons
for accelerating, rotating and charged Plebanski–Demianski black
hole have been discussed in \cite{De}. Zhang et al \cite{Zhang}
have studied the accelerating AdS black holes as the holographic
heat engines in a benchmarking scheme. Also Zhang et al \cite{Z}
have studied the thermodynamics of charged accelerating AdS black
holes and holographic heat engines. Recently Jafarzade et al
\cite{Jafar} have investigated the thermodynamics and heat engine
phenomena of charged rotating accelerating AdS black holes.
Motivated by the above works, here we'll study the thermodynamics,
$P$-$V$ criticality, stability, Joule-Thomson expansion and heat
engine for more general class of accelerating, rotating and
charged Plebanski-Demianski (PD) black hole in AdS system. In
section II, we write the general class of accelerating, rotating
and charged AdS Plebanski-Demianski black hole metric. Then we
obtain the thermodynamic quantities, critical point, specific
heat, stability and Joule-Thomson expansion. In section III, we
study the phenomena of heat engine for PD black hole and study the
Carnot cycle, Rankine cycle, work done and their efficiency.
Finally, we provide the result of the whole work in section IV.

\section{Thermodynamics of Plebanski-Demianski Black Hole}

\subsection{Plebanski-Demianski Black Hole Metric}

Plebanski and Demianski \cite{PD} have presented a large class of
Einstein-Maxwell electro-vacuum (algebraic type $D$) solutions,
which includes Kerr-Newman black hole, Taub-NUT spacetime,
(anti-)de Sitter (AdS) metric and their arbitrary combinations.
The general class of accelerating, rotating and charged
Plebanski-Demianski (PD) black hole metric in AdS system is given
by \cite{De,Griff1,Griff2,Griff3,Pod1,Pod2}
\begin{eqnarray*}
ds^{2}=\frac{1}{\Omega^{2}}\left[-\frac{{\cal
Q}}{\rho^{2}}\left\{dt-\left(a~sin^{2}\theta+4l
sin^{2}\frac{\theta}{2}\right)d\phi
\right\}^{2}+\frac{\rho^{2}}{{\cal Q}}~dr^{2} \right.
\end{eqnarray*}
\begin{equation}
\left. +\frac{{\cal
P}}{\rho^{2}}~\left\{adt+(r^{2}+(a+l)^{2})d\phi
\right\}^{2}+\frac{\rho^{2}}{{\cal P}}~sin^{2}\theta d\theta^{2}
\right]
\end{equation}
where $\Omega,~\rho^{2},~{\cal Q}$ and ${\cal P}$ are given by
\begin{equation}
\Omega=1-\frac{\alpha}{\omega}~(l+a~cos\theta)r~,
\end{equation}
\begin{equation}
\rho^{2}=r^{2}+(l+a~cos\theta)^{2}~,
\end{equation}
\begin{equation}
{\cal Q}=(\omega^{2}k+e^{2}+g^{2})-2Mr+\epsilon~ r^{2}-
\frac{2\alpha
n}{\omega}r^{3}-\left(\alpha^{2}k+\frac{\Lambda}{3}\right)r^{4},
\end{equation}
\begin{equation}
{\cal P}=(1-a_{1}~cos\theta-a_{2}~cos^{2}\theta)sin^{2}\theta~,
\end{equation}
\begin{equation}
a_{1}=\frac{2\alpha
a}{\omega}~M-\frac{4\alpha^{2}al}{\omega^{2}}~(\omega^{2}k+e^{2}+g^{2})-\frac{4}{3}al\Lambda
~,
\end{equation}
\begin{equation}
a_{2}=-\frac{\alpha^{2}a^{2}}{\omega^{2}}~(\omega^{2}k+e^{2}+g^{2})-\frac{a^{2}}{3}\Lambda~,
\end{equation}
\begin{equation}
\epsilon=\frac{\omega^{2}k}{a^{2}-l^{2}}+\frac{4\alpha
l}{\omega}M-(a^{2}+3l^{2})\left[\frac{\alpha^{2}}{\omega^{2}}(\omega^{2}k+e^{2}+g^{2})+\frac{\Lambda}{3}
\right]~,
\end{equation}
\begin{equation}
n=\frac{\omega^{2}kl}{a^{2}-l^{2}}-\frac{\alpha
(a^{2}-l^{2})}{\omega}M+l(a^{2}-l^{2})\left[\frac{\alpha^{2}}{\omega^{2}}(\omega^{2}k+e^{2}+g^{2})+\frac{\Lambda}{3}
\right]~,
\end{equation}
\begin{equation}
k=\left[1+\frac{2\alpha
l}{\omega}~M-\frac{3\alpha^{2}l^{2}}{\omega^{2}}~(e^{2}+g^{2})-l^{2}\Lambda\right]
\left[\frac{\omega^{2}}{a^{2}-l^{2}}+3\alpha^{2}l^{2}\right]^{-1}
\end{equation}
Here, $a(=J/M),~l,~e,~g,~\alpha,~\omega$ are angular momentum, NUT
parameter, electric charge, magnetic charge, acceleration
parameter and rotation parameter respectively. Also $M$ is the
mass of the PD black hole and $\Lambda$ is the cosmological
constant. In particular, the PD black hole metric can be reduced
to the following black hole metrics: (i) C-metric $(a=l=0)$
\cite{Hong1}, (ii) Kerr-Newman-Taub-NUT black hole $(\alpha=g=0)$
\cite{Mill,Bini}, (ii) Kerr-Taub-NUT black hole $(\alpha=e=g=0)$
\cite{Bini1}, (iii) Taub-NUT black hole $(\alpha=a=e=g=0)$
\cite{Iwai}, (iv) Kerr-Newman black hole $(\alpha=l=g=0)$
\cite{Lee}, (v) Kerr black hole $(\alpha=l=e=g=0)$ \cite{Bard},
(vi) Riessner-Nordstrom black hole $(\alpha=a=l=g=0)$ and (vii)
Schwarzschild black hole $(\alpha=a=l=e=g=0)$.

\subsection{Thermodynamic Quantities}

Here, we'll obtain the thermodynamic quantities for PD black hole
in AdS system. For this purpose, the cosmological constant
($\Lambda$) can be written in terms of thermodynamic pressure
($P$) as $\Lambda=-8\pi P$. The horizon radius $r_{h}$ of PD black
hole can be obtained from the following equation (putting ${\cal
Q}=0$)
\begin{equation}\label{H}
(\omega^{2}k+e^{2}+g^{2})-2Mr_{h}+\epsilon~ r_{h}^{2}-
\frac{2\alpha n}{\omega}r_{h}^{3}-\left(\alpha^{2}k-\frac{8\pi
P}{3}\right)r_{h}^{4}=0
\end{equation}

\begin{figure}
     \includegraphics[width=0.4\linewidth]{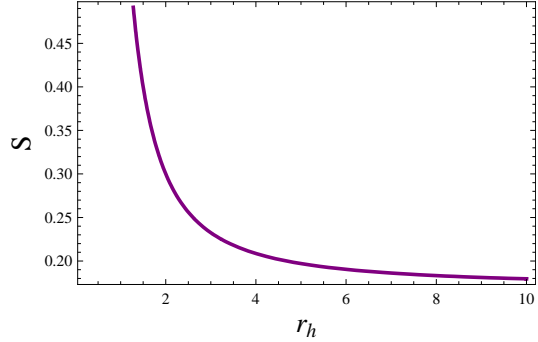}
    \caption{Figure represents the plot of entropy $S$ against PD black hole horizon radius
    $r_{h}$.}
    \label{S}
\end{figure}

\begin{figure}[!h]
     \includegraphics[width=0.4\linewidth]{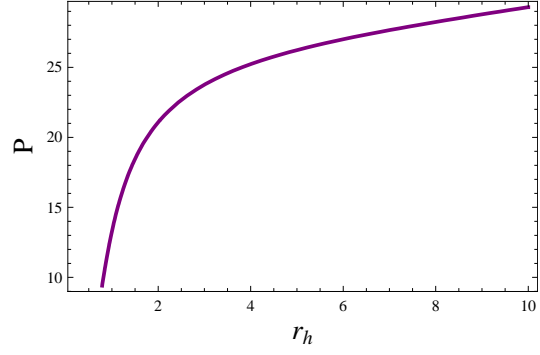}
    \caption{Figure represents the plot of pressure $P$ against PD black hole horizon radius $r_{h}$.}
    \label{P}
\end{figure}

\begin{figure}[!h]
     \includegraphics[width=0.4\linewidth]{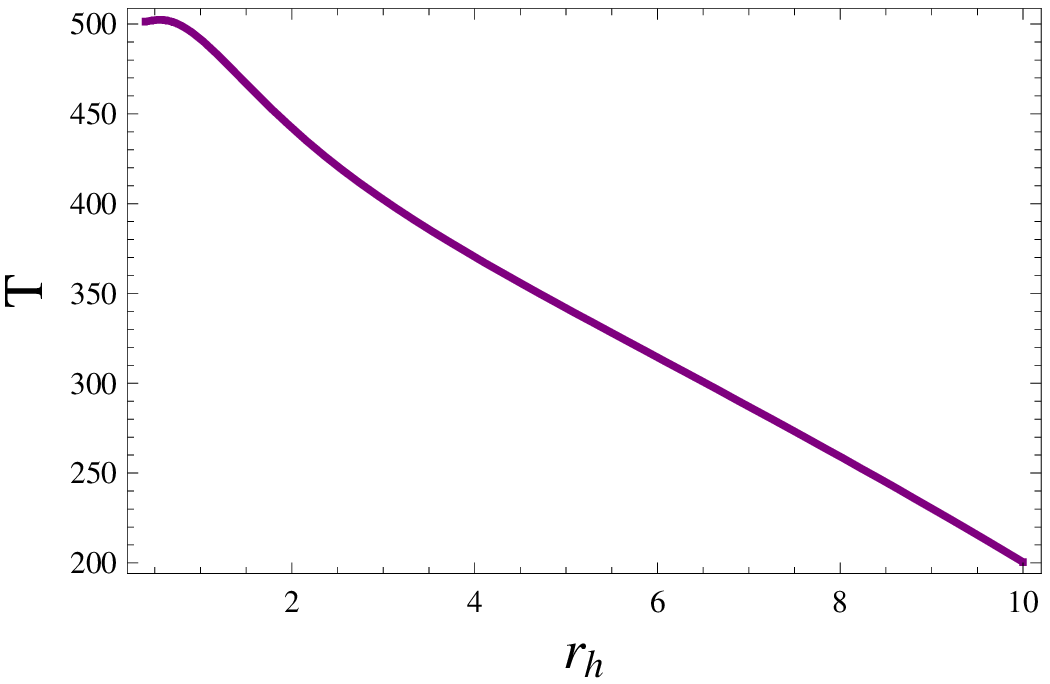}
    \caption{Figure represents the plot of temperature $T$ against PD black hole horizon radius $r_{h}$.}
    \label{T}
\end{figure}

\begin{figure}[!h]
     \includegraphics[width=0.4\linewidth]{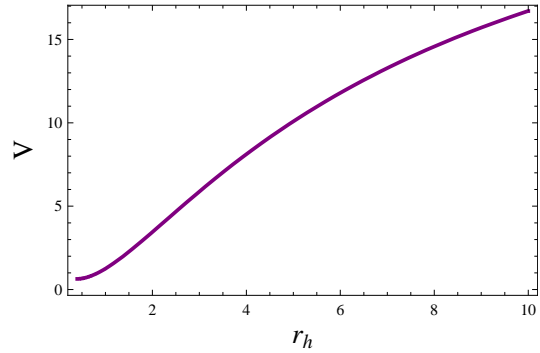}
    \caption{Figure represents the plot of volume $V$ against PD black hole horizon radius $r_{h}$.}
    \label{V}
\end{figure}

\begin{figure}
     \includegraphics[width=0.4\linewidth]{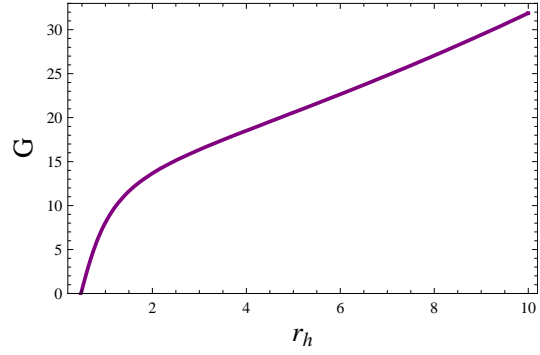}
    \caption{Figure represents the plot of Gibb's free energy $G$ against PD black hole horizon radius $r_{h}$.}
    \label{G}
\end{figure}

\begin{figure}[!h]
     \includegraphics[width=0.4\linewidth]{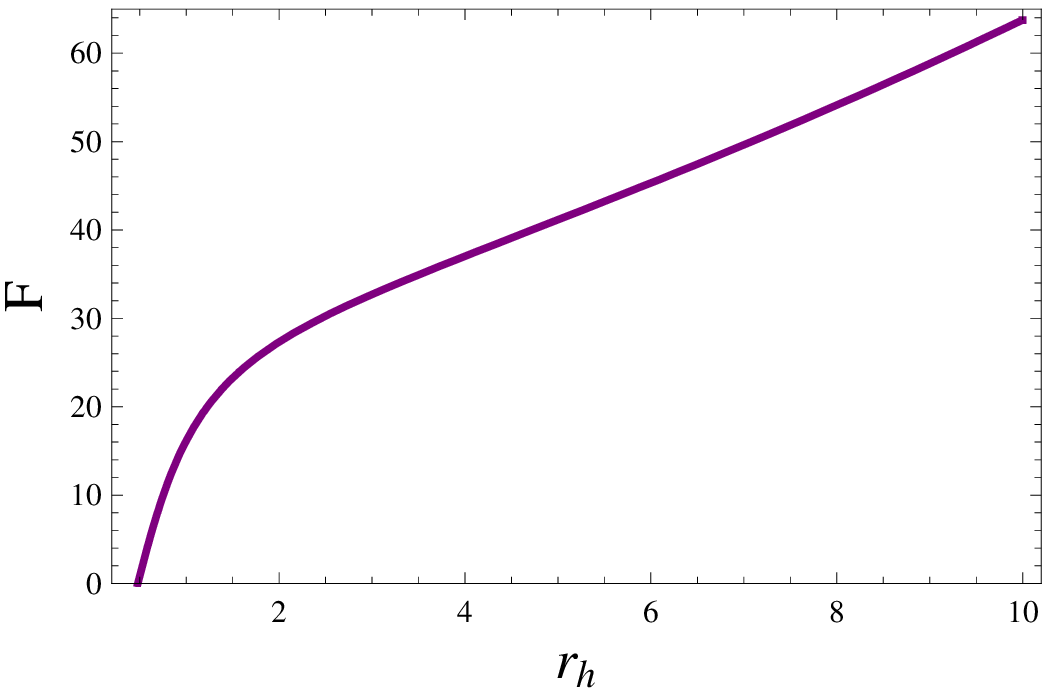}
    \caption{Figure represents the plot of Helmholtz's free energy $F$ against PD black hole horizon radius $r_{h}$.}
    \label{F}
\end{figure}

From above equation, pressure $P$ can be expressed in terms of
$r_{h}$ as
\begin{eqnarray}
&P=&\frac{3}{8 \pi \omega ^3 } \left(a^2 r_{h}^2 \alpha ^2-(\omega
-l r_{h} \alpha)^2\right) \left[2 l^4 M r_{h} \alpha ^2-2 l^3
(M+r_{h}) \alpha \omega +2 l r_{h} \left(e^2+g^2+M r_{h}\right)
\alpha\omega \right. \nonumber
\\
&&\left. +\left(e^2+g^2-2 M r_{h}+r_{h}^2\right) \omega ^2-l^2
\left(3 e^2 r_{h}^2 \alpha ^2+3 g^2 r_{h}^2 \alpha ^2+\omega
^2\right)+a^2 \left(-2 l^2 M r_{h} \alpha ^2+2 l (M+r_{h}) \alpha
\omega +\omega ^2\right)\right] \nonumber
\\
&&\times \left[-8 l^3 r_{h}^3 \alpha -3 l^4 \omega +6 l^2 r_{h}^2
\omega +r_{h}^4 \omega +a^2 \left(2 l r_{h}^3 \alpha +3 l^2 \omega
+r_{h}^2 \omega \right)\right]^{-1}
\end{eqnarray}
Now the surface area of the PD black hole in AdS system is
obtained in the form \cite{Prad}
\begin{equation}
{\cal A}=\int\int\sqrt{g_{\theta\theta}g_{\phi\phi}}~d\theta
d\phi=\frac{4\pi\omega^{2}(r_{h}^{2}+(a+l)^{2})}{(\omega-l\alpha
r_{h})^{2}-a^{2}\alpha^{2}r^{2}_{h}}
\end{equation}
So the Bekenstein-Hawking entropy \cite{B,Ba} on the horizon is
given by
\begin{equation}
S=\frac{{\cal
A}}{4}=\frac{\pi\omega^{2}(r_{h}^{2}+(a+l)^{2})}{(\omega-l\alpha
r_{h})^{2}-a^{2}\alpha^{2}r^{2}_{h}}
\end{equation}
From this, we can obtain the horizon radius in terms of entropy as
\begin{equation}
r_{h}=\frac{\omega\sqrt{f_{1}(S)}-l\alpha\omega S}{f_{2}(S)}
\end{equation}
where
\begin{equation}
f_{1}(S)=\alpha^{2}l^{2}S^{2}+\left(S-(a+l)^{2}\pi\right)f_{2}(S)
\end{equation}
and
\begin{equation}
f_{2}(S)=\alpha^{2}(a^{2}-l^{2})S+\pi\omega^{2}
\end{equation}
The volume of the PD black hole is given by
\begin{equation}
V=\left(\frac{\partial M}{\partial
P}\right)_{S,...}=\frac{4\pi\omega\left[\omega(r_{h}^{3}+6l^{2}r_{h}-6l^{4})-8\alpha
l^{3}r_{h}^{2}+a^{2}(2l\alpha r_{h}^{2}+6l^{2}\omega+\omega r_{h})
\right]}{3(3a^{2}l^{2}\alpha^{2}-3l^{4}\alpha^{2}+\omega^{2})}
\end{equation}
which can be written in terms of entropy as in the following form
\begin{equation}
V=\frac{4\pi\omega^{2}\left[6l^{2}(a^{2}-l^{2})f_{2}^{3}(S)+\omega
f_{2}(S)f_{3}(S)\left(\sqrt{f_{1}(S)}-l\alpha S
\right)+\omega^{3}\left(\sqrt{f_{1}(S)}-l\alpha S \right)^{3}
\right]}{3f_{2}^{3}(S)\left[3l^{2}\alpha^{2}(a^{2}-l^{2})+\omega^{2}
\right] }
\end{equation}
where
\begin{equation}
f_{3}(S)=a^{2}\left[f_{2}(S)+2l\alpha\left(\sqrt{f_{1}(S)}-l\alpha
S \right) \right] +2l^{2}\left[3f_{2}(S)-4l\alpha
\left(\sqrt{f_{1}(S)}-l\alpha S \right) \right]
\end{equation}
Now the surface gravity on the horizon of black hole is defined by
\begin{equation}
\kappa=\left.  \frac{1}{2\sqrt{-h}}\frac{\partial}{\partial
x^{a}}\left( \sqrt{-h}~h^{ab}\frac{\partial r}{\partial
x^{a}}\right) \right|_{r=r_{h}}
\end{equation}
where $h_{ab}$ is the second order metric constructed from the
$t$-$r$ components of the metric and $h=det(h_{ab})$. The
temperature on the horizon of the PD black hole is given by
\begin{equation}
T=\frac{\kappa}{2\pi}=\frac{[\omega-\alpha(l+a)r_{h}]}{2\pi\omega[r_{h}^{2}+(l+a)^{2}]}\left[-2M+2\epsilon~
r_{h}- \frac{6\alpha
n}{\omega}r_{h}^{2}-4\left(\alpha^{2}k-\frac{8\pi
P}{3}\right)r_{h}^{3} \right]
\end{equation}
The temperature can be written in terms of entropy as in the
following form
\begin{equation}
T=\frac{\left[f_{2}(S)+\alpha(a+l)\left(\sqrt{f_{1}(S)}-l\alpha
S\right)\right]f_{5}(S)}{3\pi f_{2}^{2}(S)f_{4}(S)}
\end{equation}
where
\begin{equation}
f_{4}(S)=(a+l)^{2}f_{2}^{2}(S)+\omega^{2}\left(\sqrt{f_{1}(S)}-l\alpha
S\right)^{2}
\end{equation}
and
\begin{eqnarray}
f_{5}(S)=3Mf_{2}^{3}(S)+\omega\left(\sqrt{f_{1}(S)}-l\alpha
S\right)\left(9n\alpha f_{2}(S)\left(\sqrt{f_{1}(S)}-l\alpha
S\right) \right. \nonumber\\
\left. +2\omega^{2}\left(\sqrt{f_{1}(S)}-l\alpha
S\right)^{2}(3k\alpha^{2}-8\pi P)-3\epsilon f_{2}^{2}(S)
\right)~~~~~~~~~~~~~~~~~~~
\end{eqnarray}
The Gibb's free energy is given by \cite{Graca}
\begin{eqnarray}
G=M-TS=\frac{(\omega^{2}+e^{2}+g^{2})f_{2}(S)}{2\omega\left(\sqrt{f_{1}(S)}-l\alpha
S\right)}+\frac{\epsilon\omega\left(\sqrt{f_{1}(S)}-l\alpha
S\right)}{2f_{2}(S)}-\frac{\alpha
n\omega\left(\sqrt{f_{1}(S)}-l\alpha S\right)^{2}}{f_{2}^{2}(S)}
\nonumber \\
-\frac{\omega^{3}}{6f_{2}^{3}(S)}(3\alpha^{2}k-8\pi
P)\left(\sqrt{f_{1}(S)}-l\alpha
S\right)^{3}-\frac{S\left[f_{2}(S)+\alpha(a+l)\left(\sqrt{f_{1}(S)}-l\alpha
S\right)\right]f_{5}(S)}{3\pi f_{2}^{2}(S)f_{4}(S)}
\end{eqnarray}
Also the Helmholtz's free energy is obtained as \cite{Graca}
\begin{eqnarray}
F=G-PV=\frac{(\omega^{2}+e^{2}+g^{2})f_{2}(S)}{2\omega\left(\sqrt{f_{1}(S)}-l\alpha
S\right)}+\frac{\epsilon\omega\left(\sqrt{f_{1}(S)}-l\alpha
S\right)}{2f_{2}(S)}-\frac{\alpha
n\omega\left(\sqrt{f_{1}(S)}-l\alpha S\right)^{2}}{f_{2}^{2}(S)}
\nonumber \\
-\frac{\omega^{3}}{6f_{2}^{3}(S)}(3\alpha^{2}k-8\pi
P)\left(\sqrt{f_{1}(S)}-l\alpha
S\right)^{3}-\frac{S\left[f_{2}(S)+\alpha(a+l)\left(\sqrt{f_{1}(S)}-l\alpha
S\right)\right]f_{5}(S)}{3\pi f_{2}^{2}(S)f_{4}(S)}  \nonumber\\
-\frac{4\pi\omega^{2}P\left[6l^{2}(a^{2}-l^{2})f_{2}^{3}(S)+\omega
f_{2}(S)\left(\sqrt{f_{1}(S)}-l\alpha S
\right)f_{3}(S)+\omega^{3}\left(\sqrt{f_{1}(S)}-l\alpha S
\right)^{3}
\right]}{3f_{2}^{3}(S)\left[3l^{2}\alpha^{2}(a^{2}-l^{2})+\omega^{2}
\right] }
\end{eqnarray}
We have drawn the entropy $S$, pressure $P$, temperature $T$,
volume $V$, Gibb's free energy $G$ and Helmholtz's free energy $F$
against PD black hole horizon radius $r_{h}$ in figures
\ref{S}-\ref{F} respectively for the parameters
$a=1.5,~l=1.2,~\omega=0.5,~\alpha=1.8,~e=1,~g=1,~M=10$. From
figure \ref{S}, we see that the entropy $S$ first sharply
decreases upto $r_{h}\approx 2$ and then slowly decreases as PD
black hole horizon radius $r_{h}$ increases. We'll choose the same
values of the parameters in all the figures. From figure \ref{P},
we see that the pressure $P$ increases as $r_{h}$ increases. From
figure \ref{T}, we see that the temperature $T$ decreases with
equal slope as $r_{h}$ increases. From figure \ref{V}, we see that
the volume $V$ increases but maintains with nearly equal slope as
$r_{h}$ grows. From figure \ref{G} and \ref{F}, we see that the
Gibb's free energy $G$ and Helmholtz's free energy $F$ first
sharply increase upto $r_{h}\approx 2$ and then slowly increase
but maintain with nearly equal slope as $r_{h}$ grows.




\subsection{$P$-$V$ Criticality}

Following \cite{Kubiz} the idea of critical behaviour of charged
AdS black holes, here we will study the critical behavior of PD
black hole. The critical points for PD black hole can be found
from the following conditions:
\begin{equation}
\left(\frac{\partial P}{\partial r_{h}}
\right)_{cr}=0,~~~~\left(\frac{\partial^{2} P}{\partial r_{h}^{2}}
\right)_{cr}=0
\end{equation}
From these conditions, we obtain the critical point $r_{cr}$ as
\begin{equation}
r_{cr}=M+2^{\frac{2}{3}}X^{-\frac{1}{3}}\left[(a+l)^{2}+M^{2}\right]+2^{-\frac{2}{3}}X^{\frac{1}{3}}
\end{equation}
where
\begin{equation}
X=M(5a^{2}+12al+4M^{2})+\sqrt{M^{2}(5a^{2}+12al+4M^{2})^{2}-16\left(
(a+l)^{2}+M^{2}\right)^{3} }
\end{equation}
At the critical point, the critical values $S_{cr}$, $P_{cr}$,
$T_{cr}$ and $V_{cr}$ are obtained as
\begin{equation}
S_{cr}=\frac{\pi\omega^{2}(r_{cr}^{2}+(a+l)^{2})}{(\omega-l\alpha
r_{cr})^{2}-a^{2}\alpha^{2}r^{2}_{cr}},
\end{equation}
\begin{equation}
P_{cr}=\frac{3\left[M-r_{cr}+\pi (a+l)^{2}T_{cr}+\pi
r^{2}_{cr}T_{cr} \right]}{8\pi
r_{cr}\left(a^{2}+6l^{2}+2r^{2}_{cr} \right) },
\end{equation}
\begin{equation}
T_{cr}=\frac{4r_{cr}^{3}-(a^{2}+6l^{2}+6r_{cr}^{2})M}{\pi\left[a^{4}+2a^{3}l+12al(l^{2}+r_{cr}^{2})+
a^{2}(7l^{2}+5r_{cr}^{2})+2(3l^{4}+r_{cr}^{4}) \right]},
\end{equation}
\begin{equation}
V_{cr}=\frac{4\pi\omega\left[\omega(r_{cr}^{3}+6l^{2}r_{cr}-6l^{4})-8\alpha
l^{3}r_{cr}^{2}+a^{2}(2l\alpha r_{cr}^{2}+6l^{2}\omega+\omega
r_{cr})\right]}{3(3a^{2}l^{2}\alpha^{2}-3l^{4}\alpha^{2}+\omega^{2})}
\end{equation}
If we choose the values of the parameters
$a=1.5,~l=1.2,~\omega=0.5,~\alpha=1.8,~e=1,~g=1,~M=10$, then we
obtain the critical point $r_{cr}=3.915$. At this critical point,
the critical values of entropy, pressure, temperature and volume
are $S_{cr}=0.21$, $P_{cr}=24.89$, $T_{cr}=382.56$ and
$V_{cr}=7.95$.

\subsection{Stability}

\begin{figure}
     \includegraphics[width=0.4\linewidth]{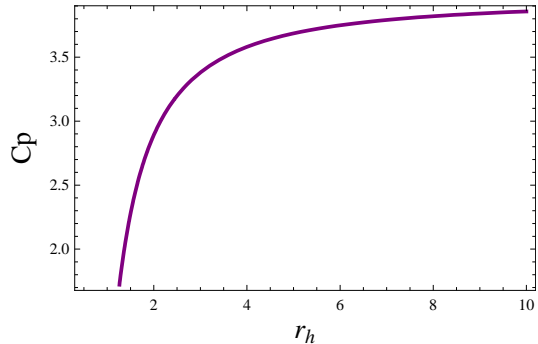}
    \caption{Figure represents the plot of specific heat capacity ${\cal C}_{P}$
    against PD black hole horizon radius $r_{h}$.}
    \label{Cp}
\end{figure}

The specific heat capacity of the black hole thermodynamical
system is the key quantity to determine the stability of the black
hole and can be written as \cite{Kubiz}
\begin{eqnarray}
{\cal C}=T\left(\frac{\partial S}{\partial T}\right)
\end{eqnarray}
If ${\cal C}\ge 0$ then the black hole will be stable and if
${\cal C}<0$ then the black hole will be unstable. If volume $V$
is constant (i.e., entropy $S$ is constant), then the specific
heat capacity ${\cal C}_{V}=0$. But for constant pressure (i.e.,
$P$ constant), we can obtain the specific heat capacity as in the
form:
\begin{eqnarray}
&&{\cal C}_{P}=\left(2 \sqrt{f_1(S)} f_2(S) \left(-(a+l) \alpha
\left(l \alpha S -\sqrt{f_1[(S)}\right)+f_2(S)\right) f_4(S)
f_5(S)\right)  \nonumber
\\
&&\times \left[(a+l) \alpha  f_2(S) f_4(S) f_5(S) f_1'(S)-2 (a+l)
\alpha f_1(S) \left(f_2(S) f_5(S) f_4'(S)+f_4(S) \left(2 f_5(S) f_2'(S)  \right.\right.\right. \nonumber
\\
&&\left.\left.\left. -f_2(S) f_5'(S)\right)\right)-2 \sqrt{f_1(S)}
\left(-2 l (a+l) S \alpha ^2 f_4(S) f_5(S) f_2'(S)+f_2^{2}(S)
\left(f_5(S) f_4'(S)-f_4(S) f_5'(S)\right)  \right.\right.
\nonumber
\\
&&\left.\left. +f_2(S) \left(-l (a+l) S \alpha ^2 f_5(S)
f_4'(S)+f_4(S) \left(f_5(S) \left(l (a+l) \alpha
^2+f_2'(S)\right)+l (a+l) S \alpha ^2
f_5'(S)\right)\right)\right)\right)]^{-1}~~~~~~
\end{eqnarray}
where dash represents derivative with respect to $S$. We have
drawn ${\cal C}_{P}$ against horizon radius $r_{h}$ in figure
\ref{Cp}. We see that the specific heat capacity ${\cal C}_{P}$
first sharply increases from some positive value upto
$r_{h}\approx 3$ and then slowly increases as $r_{h}$ grows. So
our considered PD black hole is stable in nature since from graph ${\cal C}_{P}>0$.\\

\subsection{Joule-Thomson Expansion}

\begin{figure}
     \includegraphics[width=0.4\linewidth]{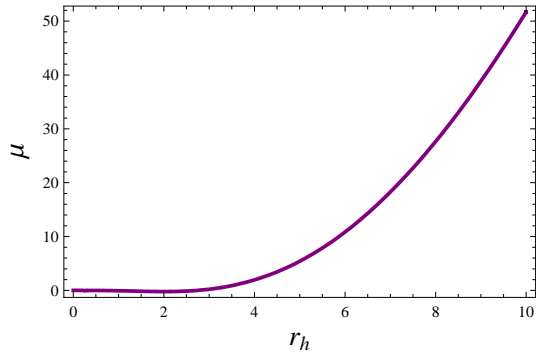}
    \caption{Figure represents the plot of Joule-Thomson coefficient $\mu$
    against PD black hole horizon radius $r_{h}$.}
    \label{Mu}
\end{figure}

Joule-Thomson expansion \cite{Win,Jo} is irreversible process,
which explains that the temperature changes from high pressure
region to low pressure region, while the enthalpy remains
constant. Since in the AdS space, the black hole mass may be
interpreted as enthalpy \cite{Kastor}, so the mass of the AdS
black hole remains constant during the Joule-Thomson expansion
process. Joule-Thomson expansion for AdS black hole has been
studied in \cite{Ok1,Ok2}. Many authors have studied the
Joule-Thomson expansion for several AdS black holes
\cite{Mo,Ch,La,Riz,Ci,Ha,Pen,Pu,Go,Ye,Nam1,Ros,Lan1,Por,G,Sade,Raj0}.
Here, we'll examine the Joule-Thomson expansion for AdS PD black
hole. The Joule-Thomson coefficient $\mu$ is the slope of the
isenthalpic curve, given by \cite{Ok1}
\begin{equation}
\mu=\left(\frac{\partial T}{\partial P}\right)_{M}
\end{equation}
which can be written as
\begin{equation}\label{mu1}
\mu=\frac{1}{{\cal C}_{P}}\left[T\left(\frac{\partial V}{\partial
T}\right)_{P}-V \right]
\end{equation}
By evaluating the sign of $\mu$, we can determine the cooling or
heating nature of the black hole. During the Joule-Thomson
expansion process, the pressure always decreases, so the change of
pressure is always negative while the change of temperature may be
positive or negative. So the temperature determines the sign of
$\mu$. If the change of temperature is positive then $\mu$ is
negative and so heating process occurs. But if the change of
temperature is negative then $\mu$ is positive and therefore
cooling process occurs. Now for AdS PD black hole, we obtain
\begin{eqnarray}
&&\mu=\frac{4 \pi  \omega ^2}{3 \left(3 l^2 \left(a^2-l^2\right)
\alpha ^2+\omega ^2\right) f_2^{4}(S)} ~ \left(-3 \left(\omega ^3
\left(-l \alpha S +\sqrt{f_1(S)}\right)^3+6 l^2
\left(a^2-l^2\right) f_2^{3}(S) \right.\right.   \nonumber
\\
&&\left. +\omega  \left(-l \alpha S +\sqrt{f_1(S)}\right) f_2(S)
f_3(S)\right) f_2'(S)+f_2(S) \left(3 \omega ^3 \left(-l \alpha S
+\sqrt{f_1(S)}\right)^2 \left(-l \alpha +\frac{f_1'(S)}{2
\sqrt{f_1(S)}}\right) \right.  \nonumber
\\
&& +\omega  f_2(S) f_3(S) \left(-l \alpha +\frac{f_1'(S)}{2
\sqrt{f_1(S)}}\right)+18 l^2 \left(a^2-l^2\right) f_2^{2}(S)
f_2'(S)+\omega  \left(-l \alpha S +\sqrt{f_1(S)}\right) f_3(S)
f_2'(S)  \nonumber
\\
&&\left. +\omega  \left(-l \alpha S +\sqrt{f_1(S)}\right) f_2(S)
f_3'(S)\right)-\left(\left(\omega ^3 \left(-l \alpha S
+\sqrt{f_1(S)}\right)^3+6 l^2 \left(a^2-l^2\right) f_2^{3}(S)
\right.\right.\nonumber
\\
&&\left. +\omega  \left(-l \alpha S +\sqrt{f_1(S)}\right) f_2(S)
f_3(S)\right) \left(-2 \left((a+l) \alpha  \left(-l \alpha S
+\sqrt{f_1(S)}\right)+f_2(S)\right) f_4(S) f_5(S) f_2'(S)  \right.
\nonumber
\\
&& +f_2(S) f_4(S) f_5(S) \left((a+l) \alpha \left(-l \alpha
+\frac{f_1'(S)}{2 \sqrt{f_1(S)}}\right)+f_2'(S)\right)-f_2(S)
\left((a+l) \alpha \left(-l \alpha S +\sqrt{f_1(S)}\right) \right.
\nonumber
\\
&&\left.\left.\left.  +f_2(S)\right) f_5(S) f_4'(S)  +f_2(S)
\left((a+l) \alpha  \left(-l \alpha S
+\sqrt{f_1(S)}\right)+f_2(S)\right) f_4(S)
f_5'(S)\right)\right)\nonumber
\\
&&\times\left.\left(\left((a+l) \alpha  \left(-l \alpha S
+\sqrt{f_1(S)}\right)+f_2(S)\right) f_4(S)
f_5(S)\right)^{-1}\right)
\end{eqnarray}
We have drawn the Joule-Thomson coefficient $\mu$ against PD black
hole horizon radius $r_{h}$ in figure \ref{Mu}. We observe that
$\mu$ first keeps nearly parallel to $r_{h}$ axis upto
$r_{h}\approx 3$ but keeps positive sign and then increases as
$r_{h}$ increases. Since $\mu>0$ throughout the expansion of
$r_{h}$, so the change of temperature is negative and therefore
cooling process occurs for PD black hole.\\

\begin{figure}
     \includegraphics[width=0.4\linewidth]{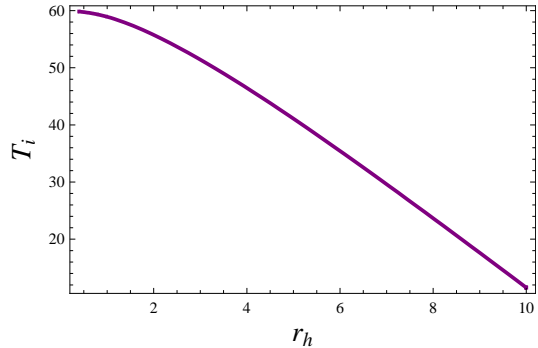}
    \caption{Figure represents the plot of inversion temperature $T_{i}$
    against PD black hole horizon radius $r_{h}$.}
    \label{Ti}
\end{figure}

For $\mu=0$, we can determine the expansion process of inversion
curve in a small infinite pressure. At the inversion temperature,
put $\mu=0$ in (\ref{mu1}) and inversion temperature is given by
\begin{equation}
T_{i}=V\left(\frac{\partial T}{\partial V}\right)_{P}
\end{equation}
So for PD black hole, we obtain the inversion temperature
\begin{eqnarray}
&T_{i}&=\left[3 l S \alpha  \omega ^3 f_1(S)-\omega ^3
f_1^{3/2}(S)-\sqrt{f_1(S)} \left(3 l^2 S^2 \alpha ^2 \omega
^3+\omega f_2(S) f_3(S)\right) \right. \nonumber
\\
&&\left.\left.+l \left(l^2 S^3 \alpha ^3 \omega ^3+\left(-6 a^2
l+6 l^3\right) f_2^{3}(S)+S \alpha  \omega  f_2(S)
f_3(S)\right)\right) \left((a+l) \alpha f_2(S) f_4(S) f_5(S)
f_1'(S) \right.\right. \nonumber
\\
&&-2 (a+l) \alpha f_1(S) \left(f_2(S) f_5(S) f_4'(S)+f_4(S)
\left(2 f_5(S) f_2'(S)-f_2(S) f_5'(S)\right)\right) \nonumber
\\
&&-2\sqrt{f_1(S)} \left(-2 l (a+l) S \alpha ^2 f_4(S) f_5(S)
f_2'(S)+f_2^{2}(S) \left(f_5(S) f_4'(S)-f_4(S) f_5'(S)\right)
\right. \nonumber
\\
&&\left.\left.+f_2(S) \left(-l (a+l) S \alpha ^2 f_5(S)
f_4'(S)+f_4(S) \left(f_5(S) \left(l (a+l) \alpha
^2+f_2'(S)\right)+l (a+l) S \alpha ^2
f_5'(S)\right)\right)\right)\right]  \nonumber
\\
&&\times\left[3 \pi \omega f_2^{2}(S) f_4^{2}(S) \left(-f_2(S)
\left(3 l^2 S^2 \alpha ^2 \omega ^2+f_2(S) f_3(S)\right) f_1'(S)+6
\omega ^2 f_1^{2}(S) f_2'(S) \right.\right.  \nonumber
\\
&&+6 l \alpha  \omega ^2 f_1^{3/2}(S) \left(f_2(S)-3 S
f_2'(S)\right)+f_1(S) \left(18 l^2 S^2 \alpha ^2 \omega ^2
f_2'(S)+f_2(S) \left(-12 l^2 S \alpha ^2 \omega ^2-3 \omega ^2
f_1'(S) \right.\right.  \nonumber
\\
&&\left.\left.+4 f_3(S) f_2'(S)\right)-2 f_2^{2}(S)
f_3'(S)\right)+2 l \alpha \sqrt{f_1(S)} \left(-3 l^2 S^3 \alpha ^2
\omega ^2 f_2'(S)+S f_2(S) \left(3 l^2 S \alpha ^2 \omega ^2+3
\omega ^2 f_1'(S) \right.\right.   \nonumber
\\
&&\left.\left.\left.\left.-2 f_3(S) f_2'(S)\right)+f_2^{2}(S)
\left(f_3(S)+S f_3'(S)\right)\right)\right)\right]^{-1}
\end{eqnarray}
We have drawn the inversion temperature $T_{i}$ against PD black
hole horizon radius $r_{h}$ in figure \ref{Ti}. We observe that
$T_{i}$ decreases slowly as $r_{h}$ increases but slopes of the
curve al all points are almost same throughout the evolution of
$r_{h}$.


\section{\bf{Heat Engine}}

A heat engine is a thermodynamic system that converts thermal
energy (or heat) and chemical energy to mechanical energy to do
mechanical work. So physically, a heat engine carries heat from
hot reservoir in which part of the heat converts into the physical
works while the remaining part of the heat is transferred to cold
reservoir. The working substance in a black hole heat engine is
thought to be the black hole fluid or black hole molecules. The
heat engine brings a working substance from a higher state
temperature to a lower state temperature. Then the working
substance generates work while transferring heat to the cold
reservoir until it reaches a low temperature state. During this
process in the heat engine, some of the thermal energy is
converted into work where the working substance has non-zero heat
capacity. Therefore, the heat engine operates in a cyclic manner
by adding energy (heat) in one part of the cycle and using that
energy to do work in another part of the cycle. In the following
subsections, we'll study the Carnot cycle and Rankine cycle of the
heat engine for AdS PD black hole.

\subsection{Carnot Cycle}

Carnot cycle is theoretical ideal thermodynamic cycle which was
proposed by N. L. S. Carnot in 1824. A Carnot heat engine is a
classical thermodynamic engine that operates on the Carnot cycle
which can be achieved during the conversion of heat into work.
There are two heat reservoirs (hot and cold) forming part of the
heat engine of the Carnot cycle. Here we assume, $T_{H}$ and
$T_{C}$ are temperatures of hot and cold reservoirs respectively.
These are included upper and lower isothermal processes with two
adiabatic processes. The $P$-$V$ diagram has been shown in Ref
\cite{John} for Carnot heat engine which forms a closed path. From
the diagram, it was shown that along the upper isotherm process,
the heat flows are generated from 1 to 2 and which is given by
\cite{John}
\begin{equation}
Q_{H}=T_{H}\triangle S_{1\rightarrow 2}=T_{H}(S_{2}-S_{1})
\end{equation}
and the exhausted heat produced from 3 to 4 along lower isothermal
process is given by \cite{John}
\begin{equation}
Q_{C}=T_{C}\triangle S_{3\rightarrow 4}=T_{C}(S_{3}-S_{4})
\end{equation}
Here PD black hole entropies $S_{i}$'s are related to volumes
$V_{i}$'s satisfying as
\begin{equation}\label{37}
V_{i}=\frac{4\pi\omega^{2}\left[6l^{2}(a^{2}-l^{2})f_{2}^{3}(S_{i})+\omega
f_{2}(S_{i})f_{3}(S_{i})\left(\sqrt{f_{1}(S_{i})}-l\alpha S_{i}
\right)+\omega^{3}\left(\sqrt{f_{1}(S_{i})}-l\alpha S_{i}
\right)^{3}
\right]}{3f_{2}^{3}(S_{i})\left[3l^{2}\alpha^{2}(a^{2}-l^{2})+\omega^{2}
\right] }~,~i=1,2,3,4,
\end{equation}
where $f_{1}(S_{i})$, $f_{2}(S_{i})$ and $f_{3}(S_{i})$ can be
calculated from the following relations
\begin{equation}
f_{1}(S_{i})=\alpha^{2}l^{2}S_{i}^{2}+\left(S_{i}-(a+l)^{2}\pi\right)f_{2}(S_{i})~,~i=1,2,3,4,
\end{equation}
\begin{equation}
f_{2}(S_{i})=\alpha^{2}(a^{2}-l^{2})S_{i}+\pi\omega^{2}~,~i=1,2,3,4
\end{equation}
and
\begin{equation}
f_{3}(S_{i})=a^{2}\left[f_{2}(S_{i})+2l\alpha\left(\sqrt{f_{1}(S_{i})}-l\alpha
S_{i} \right) \right] +2l^{2}\left[3f_{2}(S_{i})-4l\alpha
\left(\sqrt{f_{1}(S_{i})}-l\alpha S_{i} \right)
\right]~,~i=1,2,3,4.
\end{equation}
Also the heat engine flow has been shown in Ref \cite{John}. The
total mechanical work done by the heat engine is the difference of
the amount of heat energies between upper and lower isotherm
processes as
\begin{equation}
W=Q_{H}-Q_{C}
\end{equation}
A central quantity, the efficiency of a Carnot heat engine is
defined by the ratio of total mechanical work and the amount of
heat energy along the upper isotherm process and is given by
\begin{equation}
\eta_{_{Car}}=\frac{W}{Q_{H}}=1-\frac{Q_{C}}{Q_{H}}
\end{equation}
Since for Carnot cycle, $V_{4}=V_{1}$ and $V_{3}=V_{2}$, so we
have the maximum efficiency for Carnot cycle and is given by
\begin{equation}
\left.(\eta_{_{Car}})_{_{max}}=1-\frac{T_{C}(S_{3}-S_{4})}{T_{H}(S_{2}-S_{1})}
\right|_{V_{4}=V_{1},V_{3}=V_{2}}
\end{equation}
which simplifies to the following form
\begin{equation}
(\eta_{_{Car}})_{_{max}}=1-\frac{T_{C}}{T_{H}}
\end{equation}
Since $T_{H}>T_{C}$, so we have $0<(\eta_{_{Car}})_{_{max}}<1$. It
should be noted that the $(\eta_{_{Car}})_{_{max}}$ is the maximum
efficiency of all possible cycles between higher temperature
$T_{H}$ and lower temperature $T_{C}$. Since we know that the
Stirling cycle consists of two isothermal processes and two
isochores processes, so this maximally efficient Carnot engine is
also Stirling engine. From the figure \ref{T}, if we choose
$T_{C}=250$ and $T_{H}=400$ then the maximum efficiency for Carnot
cycle is obtained as
$(\eta_{_{Car}})_{_{max}}=0.375$.\\

Now we discuss a new engine of black hole which consists of two
isobars and two isochores/adiabats described in Ref \cite{John},
where the heat flows show along the top and bottom. The total work
done along the isobars is given by
\begin{eqnarray}
W=\triangle P_{4\rightarrow 1}~\triangle V_{1\rightarrow
2}=(P_{1}-P_{4})(V_{2}-V_{1})
\end{eqnarray}
where $V_{1}$ and $V_{2}$ are described in (\ref{37}). The upper
isobar gives the net inflow of heat which is given by
\begin{eqnarray}
&Q_{H}&=\int_{T_{1}}^{T_{2}} {\cal C}_{P}(P_{1},T)dT  \nonumber\\
&&=\left.\int_{S_{1}}^{S_{2}}\frac{\left[f_{2}(S)+\alpha(a+l)\left(\sqrt{f_{1}(S)}-l\alpha
S\right)\right]f_{5}(S)}{3\pi
f_{2}^{2}(S)f_{4}(S)}\right|_{P=P_{1}}dS
\end{eqnarray}
The lower isobar gives the exhaust of heat which is given by
\begin{eqnarray}
&Q_{C}&=\int_{T_{3}}^{T_{4}} {\cal C}_{P}(P_{4},T)dT    \nonumber\\
&&=\left.\int_{S_{3}}^{S_{4}}\frac{\left[f_{2}(S)+\alpha(a+l)\left(\sqrt{f_{1}(S)}-l\alpha
S\right)\right]f_{5}(S)}{3\pi
f_{2}^{2}(S)f_{4}(S)}\right|_{P=P_{4}}dS
\end{eqnarray}
The thermal efficiency for the new heat engine is described by
\begin{eqnarray}
&\eta_{_{New}}&=\frac{W}{Q_{H}}=\frac{(P_{1}-P_{4})(V_{2}-V_{1})}{Q_{H}}
\nonumber\\
&&=(P_{1}-P_{4})(V_{2}-V_{1})\left[\left.\int_{S_{1}}^{S_{2}}\frac{\left[f_{2}(S)+\alpha(a+l)
\left(\sqrt{f_{1}(S)}-l\alpha S\right)\right]f_{5}(S)}{3\pi
f_{2}^{2}(S)f_{4}(S)}\right|_{P=P_{1}}dS\right]^{-1}
\end{eqnarray}
Since the above integration is very complicated, so we may find
the value of the efficiency of the new engine for particular
values of the parameters, say
$a=1.5,~l=1.2,~\omega=0.5,~\alpha=1.8,~e=1,~g=1,~M=10$. If we
choose
$S_{1}=0.25,~S_{2}=0.40,~P_{1}=15,~P_{4}=25,~V_{1}=5,~V_{2}=15$
from the figures \ref{S}, \ref{P} and \ref{V}, then the value of
the thermal efficiency of the new engine is obtained as
$\eta_{_{New}}=0.907$.

\subsection{Rankine Cycle}

A Rankine cycle \cite{Wei00} is an idealized thermodynamic cycle
of a heat engine, which converts heat into mechanical work for
undergoing phase change. The Rankine cycle for black hole heat
engine is shown in Ref \cite{Wei1}. From the Ref \cite{Wei1}, we
see that the working substance starts from $A$ to $B$ with
increasing temperature and adiabatic pressure. Next the working
substance follows from $B$ to $E$ and within these states a phase
transition occurs from $C$ to $D$ with constant temperature. Then
the working substance reduces the temperature from $E$ to $F$ and
returns to $A$ by reducing its volume. For constant pressure $P$,
we have $dP=0$. From the first law of the black hole
thermodynamics $dH_{P}=TdS$ for constant pressure, we have the
enthalpy function $H_{P}(S)=\int TdS$. Now according to the
formalism of Wei et al \cite{Wei00,Wei1}, the efficiency for
Rankine cycle for black hole heat engine can be expressed as
\begin{equation}\label{etaR}
\eta_{_{Ran}}=1-\frac{T_{A}(S_{F}-S_{A})}{H_{P_{_{B}}}(S_{F})-H_{P_{_{B}}}(S_{A})}
=1-\frac{T_{1}(S_{3}-S_{1})}{H_{P_{_{2}}}(S_{3})-H_{P_{_{2}}}(S_{1})}
\end{equation}
where the subscripts $A,~B,~F$ are denoted by $1,~2,~3$
respectively. Here $T_{1}$ and $H _{P_{_{2}}}$ for PD black hole
are given by
\begin{equation}
T_{1}=\frac{\left[f_{2}(S_{1})+\alpha(a+l)\left(\sqrt{f_{1}(S_{1})}-l\alpha
S_{1}\right)\right]f_{5}(S_{1})}{3\pi
f_{2}^{2}(S_{1})f_{4}(S_{1})}
\end{equation}
and
\begin{equation}
H_{P_{_{2}}}(S_{i})=\left.\int_{S_{0}}^{S_{i}}
\frac{\left[f_{2}(S)+\alpha(a+l)\left(\sqrt{f_{1}(S)}-l\alpha
S\right)\right]f_{5}(S)}{3\pi
f_{2}^{2}(S)f_{4}(S)}\right|_{P=P_{2}}dS~,~i=1,3
\end{equation}
So equation (\ref{etaR}) can be expressed in the following form:
\begin{eqnarray}
&\eta_{_{Ran}}&=1-\frac{(S_{3}-S_{1})\left[f_{2}(S_{1})+\alpha(a+l)\left(\sqrt{f_{1}(S_{1})}-l\alpha
S_{1}\right)\right]f_{5}(S_{1})}{3\pi
f_{2}^{2}(S_{1})f_{4}(S_{1})}  \nonumber
\\
&&\times
\left[\left.\int_{S_{1}}^{S_{3}}\frac{\left[f_{2}(S)+\alpha(a+l)
\left(\sqrt{f_{1}(S)}-l\alpha S\right)\right]f_{5}(S)}{3\pi
f_{2}^{2}(S)f_{4}(S)}\right|_{P=P_{2}}dS\right]^{-1}
\end{eqnarray}
Since the above integration is very complicated, so we may find
the value of the efficiency of the Rankine cycle for particular
values of the parameters, say
$a=1.5,~l=1.2,~\omega=0.5,~\alpha=1.8,~e=1,~g=1,~M=10$. If we
choose $S_{1}=0.25,~S_{3}=0.45,~P_{2}=15$ from the figures \ref{S}
and \ref{P}, then the value of the efficiency of the rankine cycle
is obtained as $\eta_{_{Ran}}=0.968$.

\section{Discussions and Concluding Remarks}

We have assumed the general class of accelerating, rotating and
charged Plebanski-Demianski (PD) black holes in presence of
cosmological constant, which includes the Kerr-Newman rotating
black hole and the Taub-NUT spacetime. We have assumed that the
thermodynamical pressure ($P$) may be described as the negative
cosmological constant ($\Lambda<0$) by the relation $\Lambda=-8\pi
P$ and so the black hole may be formed in anti-de Sitter (AdS) PD
black hole. Using the horizon radius ($r_{h}$), the thermodynamic
quantities like surface area ($\kappa$), entropy ($S$), volume
($V$), temperature ($T$), Gibb's free energy ($G$) and Helmholtz's
free energy ($F$) of the AdS PD black hole have been obtained due
to the thermodynamic system. We have drawn the entropy $S$,
pressure $P$, temperature $T$, volume $V$, Gibb's free energy $G$
and Helmholtz's free energy $F$ against PD black hole horizon
radius $r_{h}$ in figures \ref{S}-\ref{F} respectively for the
parameters $a=1.5,~l=1.2,~\omega=0.5,~\alpha=1.8,~e=1,~g=1,~M=10$.
From figure \ref{S}, we have observed that the entropy $S$ first
sharply decreases upto $r_{h}\approx 2$ and then slowly decreases
as PD black hole horizon radius $r_{h}$ increases. From figure
\ref{P}, we have seen that the pressure $P$ increases as $r_{h}$
increases. From figure \ref{T}, we have seen that the temperature
$T$ decreases with equal slope as $r_{h}$ increases. From figure
\ref{V}, we have observed that the volume $V$ increases but
maintains with nearly equal slope as $r_{h}$ grows. From figure
\ref{G} and \ref{F}, we have seen that the Gibb's free energy $G$
and Helmholtz's free energy $F$ first sharply increase upto
$r_{h}\approx 2$ and then slowly increase but maintain with nearly
equal slope as $r_{h}$ grows.\\

Next we found the critical point and corresponding critical
entropy, critical pressure, critical temperature and critical
volume for AdS PD black hole. In particular, for the chosen the
values of the parameters
$a=1.5,~l=1.2,~\omega=0.5,~\alpha=1.8,~e=1,~g=1,~M=10$, we have
obtained the critical point $r_{cr}=3.915$ and the corresponding
critical values of entropy, pressure, temperature and volume are
$S_{cr}=0.21$, $P_{cr}=24.89$, $T_{cr}=382.56$ and $V_{cr}=7.95$.
We have drawn ${\cal C}_{P}$ against horizon radius $r_{h}$ in
figure \ref{Cp}. We have seen that the specific heat capacity
${\cal C}_{P}$ first sharply increases from some positive value
upto $r_{h}\approx 3$ and then slowly increases as $r_{h}$ grows.
Due to the study of specific heat capacity, we have obtained
${\cal C}_{V}=0$ and  from the graph, we have obtained ${\cal
C}_{P}>0$. From this result, we have concluded that the AdS PD
black hole may be stable. We have also examined the Joule-Thomson
expansion of PD black hole and by evaluating the sign of
Joule-Thomson coefficient $\mu$, we have determined the heating
and cooling nature of PD black hole. We have drawn the
Joule-Thomson coefficient $\mu$ against PD black hole horizon
radius $r_{h}$ in figure \ref{Mu}. We have observed that $\mu$
first keeps nearly parallel to $r_{h}$ axis upto $r_{h}\approx 3$
but keeps positive sign and then increases as $r_{h}$ increases.
Since $\mu>0$ throughout the expansion of $r_{h}$, so the change
of temperature is negative and therefore cooling process occurs
for PD black hole. Putting $\mu=0$, we have obtained the inversion
temperature $T_{i}$. We have drawn the inversion temperature
$T_{i}$ against PD black hole horizon radius $r_{h}$ in figure
\ref{Ti}. We have observed that $T_{i}$ decreases slowly as
$r_{h}$ increases but slopes of the curve al all points are almost
same throughout the evolution of $r_{h}$.\\

Next we have studied the heat engine phenomena for AdS PD black
hole. We have analyzed the heat flows from upper and lower
isotherms process. For the heat engine in Carnot cycle, we have
calculated the work done and maximum efficiency. If we choose
$T_{C}=250$ and $T_{H}=400$ then the maximum efficiency for Carnot
cycle is obtained as $(\eta_{_{Car}})_{_{max}}=0.375$. Also for a
new engine, we have assumed another cycle which consists of two
isobars and two isochores. Then we have calculated the net inflow
of heat in upper isobar, work done and its efficiency of the new
heat engine for this cycle. In particular, for
$S_{1}=0.25,~S_{2}=0.40,~P_{1}=15,~P_{4}=25,~V_{1}=5,~V_{2}=15$
the value of the thermal efficiency of the new engine is obtained
as $\eta_{_{New}}=0.907$. Finally, we have analyzed the efficiency
for the Rankine cycle in PD black hole heat engine. In particular,
for $S_{1}=0.25,~S_{3}=0.45,~P_{2}=15$ the value of the efficiency
of the rankine cycle is obtained as $\eta_{_{Ran}}=0.968$.


\end{document}